\documentclass[12pt]{article}
\usepackage{amsfonts}
\usepackage{epsfig}
\usepackage{psfig}
\usepackage{graphicx}
\usepackage{amsmath}
\begin{document}

\title{Maxwell stress tensor and the Casimir effect}
\author{C. Farina, J. J. Passos Sobrinho and A. C. Tort \\
Instituto de F\'{\i}sica\break \\
Universidade Federal do Rio de Janeiro\break \\
Cidade Universit\'aria - Ilha do Fund\~ao - Caixa Postal 68528\\
21945-970 Rio de Janeiro RJ, Brasil.}
\date{\today}
\maketitle

\begin{abstract}
We evaluate the quantum correlators associated with the Maxwell field vacuum
distorted by the presence of plane parallel material surfaces.
Regularization is performed through the generalized zeta funtion technique.
Results are applied to a local analysis of the atractive and repulsive
Casimir effect through Maxwell stress tensor. Surface divergences are shown
to cancel out when stresses on both sides of the material surface are taken
into account. Since an atom can be considered as a probe of the local
distortion of the quantum vacuum, Casimir-Polder interactions between atoms
and material surfaces are also considered.
\end{abstract}

\noindent PACS: 11. 10. -z; 12. 20. -m

\vfill

\noindent $^{\dagger }$ {e-mail: farina@if.ufrj.br}

\noindent $^{\ddagger }$ {e-mail:tort@if.ufrj.br}\newpage

\section{Introduction}

According to Casimir \cite{Casimir}, the (macroscopically) observable vacuum
energy of a quantum field is the regularized difference between the zero
point energies with and without the external conditions demanded by the
particular physical situation at hand. In the case of the quantized
electromagnetic field confined between two infinite parallel conducting
plates separated by a distance $a$, Casimir's conception of the vacuum
energy leads to a force per
unit area between the plates given by 
\begin{equation}
\frac{F}{A}=-\frac{\pi \hbar c}{240\,a^{4}}.
\end{equation}
Until 1997 only one experiment involving Casimir's original setup had been
performed \cite{Sparnaay}. Recently, however, the experimental observation
of this tiny force with metallic surfaces was significantly improved by the
experiments due to Lamoreaux and to Mohideen and Roy \cite{Lamo}.The concept
of an observable vacuum energy can be extended to all quantum fields and
several types of boundary conditions and/or applied external fields. A
review of all these Casimir effects can be found in, for example,
Mostepanenko and Trunov or in Plunien \textit{et al. }\cite{MosteTrunov}.

The local approach to the electromagnetic Casimir effect was initiated by
Brown and Maclay who calculated the renormalized stress energy tensor
between two parallel perfectly conducting plates by means of Green functions
techniques \cite{Brown&Maclay69}. An interesting approach to the standard
Casimir effect is the one due to Gonzales \cite{Gonzales}. This author pointed out that the apparently non-objectionable definition of
the vacuum energy given above could easily lead to conceptual errors and
stressed the fact that in any Casimir interaction calculation, contributions
from both sides of the material surfaces involved must be taken into
account. This is so because the vacuum pressure always pushes the material
surfaces involved, therefore, the repulsiveness or atractiveness
of the Casimir force depends on the discontinuity of the relevant component
of the quantized Maxwell stress tensor at the location of the surface.

The purpose of this paper is to pursue this line of reasoning by analyzing
the stresses on parallel material surfaces due to the vacuum distortions
caused by the presence of these surfaces through the quantized Maxwell
stress tensor. Though the approach chosen here has many points in common
with Ref. \cite{Gonzales} cited above, it is a different alternative in the sense that
it relies on objects known as correlators, which are vacuum expectation
values of products of field components taken at the same point in space and
time. These correlators contain all the information we need on the local
behavior of the vacuum expectation values of elements of Maxwell stress
tensor, in particular, their behavior near both sides of the material
surface in question, a feature crucial to the obtention of the correct
result. The local behavior of the Maxwell tensor, or of the relativistic
symmetrical stress-energy tensor, is extremely important because as shown
by, for example, Deutsch and Candelas \cite{Deutsch79} with the help of
Green functions technique, as we approach the boundaries we find strong
divergencies that cannot be removed by renormalization. Besides reviewing
the obtention of the electromagnetic Casimir for the standard case of two
perfectly conducting parallel plates, we will also consider a pair of
parallel plates, one of them perfectly permeable. This setup was first
proposed by Boyer \cite{Boyer74} who analyzed them from the viewpoint of
random electrodymanics and it is the simplest example of a repulsive Casimir
force. For both cases, \ the conducting plate and Boyer's setup we also
construct the symmetrical stress tensor. 

It is well-known that an atom can be attracted or
repelled by a material surface, thus probing locally the vacuum distortions,
for these reasons we take advantage of the knowledge of the above correlators and include a derivation of the Casimir-Polder
interaction between an atom and a material surface. We will employ gaussian
units and set $c=\hbar =1$.

\section{Maxwell stress tensor and the electromagnetic field correlators}

Our aim in this section is to obtain an expression for the quantum version
of the electromagnetic force per unit area that acts on plane material
surface. Material surface here means a perfectly conducting square surface($%
\epsilon \rightarrow \infty $) or a perfectly permeable one ($\mu
\rightarrow \infty )$ whose linear dimension $L$ is much larger than others
relevant dimensions envolved such as the distance between two of those
surfaces. The physical interaction between any of the two types of surfaces
considered here and the vacuum electromagnetic field is mimicked by the
imposition of appropriate boundary conditions on the electromagnetic field
on the location of the material surface. The Cartesian components of the
Maxwell stress tensor in Gaussian units are given by \cite{Jackson3rd}
\begin{equation}
T_{ij}=\frac{1}{4\pi }\left( E_{i}\,E_{j}-\frac{1}{2}\,\delta _{ij}\mathbf{E}%
^{2}+B_{i}\,B_{j}-\frac{1}{2}\,\delta _{ij}\mathbf{B}^{2}\right)
\label{MaxwellStressTensor}
\end{equation}
where $i,j=x,y,z$. Suppose that the material surface is placed perpendicularly to the $\mathcal{OZ}$ axis. Upon
quantizing the electromagnetic field we can write the quantum version of (\ref{MaxwellStressTensor}). For instance,
\begin{equation}
\left\langle \hat{T}_{zz}\right\rangle _{0}=\frac{1}{8\pi }\left[
\left\langle \hat{E}_{z}^{2}\right\rangle _{0}-\left\langle \hat{E}%
_{\parallel }^{2}\right\rangle _{0}+\left\langle \hat{B}_{z}^{2}\right%
\rangle _{0}-\left\langle \hat{B}_{\parallel }^{2}\right\rangle _{0}\right] ,
\label{MainEq}
\end{equation}
where $\hat{E}_{\parallel }^{2}=\hat{E}_{x}^{2}+\hat{E}_{y}^{2}$ and $\hat{B}%
_{\parallel }^{2}=\hat{B}_{x}^{2}+\hat{B}_{y}^{2}$; $\left\langle \hat{O}%
\right\rangle _{0}\equiv \left\langle 0\right| \hat{O}\left| 0\right\rangle $
denotes a vacuum expectation value. The other Cartesian components of this
tensor can be obtained in an analogous way. The quantum macroscopic force$%
\left\langle \mathbf{\hat{F}}\right\rangle _{0}$ on the material surface can
be evaluated by integrating the quantum version of the classical result\cite
{Jackson3rd}
\begin{equation}
\mathbf{F=}\oint_{\partial \mathcal{R}}\mathbf{\tilde{T}\cdot \hat{n}}\,da,
\label{ForceonV}
\end{equation}
were $\mathbf{\hat{n}}$ is outwards normal at $\partial \mathcal{R}$ and $%
\mathcal{R}$ is any region containing the material surface. Classically, (\ref{ForceonV}) can be obtained by integrating the Lorentz force per unit
volume acting on charge and current distributions and eliminating the sources in favor of the fields. From a quantum point of view, we see that
the problem of evaluating the pressure the material surface due to the
distorted zero point oscillations of the electromagnetic field is reduced to
the evaluation of the vacuum expectation value of the quantum operators $\hat{E}_{i}\left( \mathbf{r},t\right) \hat{E}_{j}\left( \mathbf{r},t\right) $, $\hat{B}_{i}\left( \mathbf{r},t\right) \hat{B}_{j}\left( \mathbf{r},t\right) ,$ and $\hat{E}_{i}\left( \mathbf{r},t\right) \hat{B}_{j}\left( 
\mathbf{r},t\right) $. The evaluation of these correlators depends on the
specific choice of the boundary conditions. A regularization recipe will
also be necessary, for these objects are mathematically ill-defined. For the
setup envolving conducting plates these correlators were evaluated by
L\"{u}tken and Ravndal \cite{Lü&Ravndal85}, see also \cite{Barton90}. They
can be also obtained from the coincidence limit of the photon propagator
between conducting plates evaluated by Bordag \textit{et al} \cite{Bordag85}. In the next section we will display a method of evaluating these
correlators by means of analitycal continuation techniques \cite{JPA99}
similar to the ones employed by L\"{u}tken e Ravndal \cite{Lü&Ravndal85}. We will also show how to obtain the
corresponding results for another unusual but intersting setup \cite{JPA99}.

\section{Correlators for Casimir's setup}

Consider an experimental setup consisting in two infinite perfectly
conducting parallel plates ($\epsilon \rightarrow \infty $) kept at a fixed
distance $a$ from each other. We will choose the coordinates axis in such a
way that the $\mathcal{OZ}$ direction is perpendicular to the plates. One of
the plates will be placed at $z=0$ and the othrt one at $z=a$. The field
must satisfy the following boundary conditions on the plates: the tangential
components $E_{x}$ e $E_{y}$ of the electric field and the normal component $%
B_{z}$ of the magnetic field must be zero on the plates. Since there are no
real charges or currents it will be convenient to work in the Coulomb gauge
in which $\mathbf{\nabla }\cdot \mathbf{A}(\mathbf{r},t)=0$, and $\Phi =0$,
thus $\mathbf{E}(\mathbf{r},t)=-\partial \mathbf{A}(\mathbf{r},t)/\partial t$
and $\mathbf{B}(\mathbf{r},t)=\mathbf{\nabla }\times \mathbf{A}\left( 
\mathbf{r,t}\right) $. These physical boundary conditions combined with the
choice of gauge allow us to rewrite the boundary conditions in terms of the
components of the vector potential $\mathbf{A}(\mathbf{r},t)$ in the
following way: at $\ z=0$ we will have, 
\begin{equation}
A_{x}(x,y,0,t)=0\,;\;\;\;\;A_{y}(x,y,0,t)=0\,;\;\;\;\;{\frac{\partial }{%
\partial z}}A_{z}(x,y,0,t)=0\,,
\end{equation}
and at $z=a$ , 
\begin{equation}
A_{x}(x,y,a,t)=0\,;\;\;\;\;A_{y}(x,y,a,t)=0\,;\;\;\;\;{\frac{\partial }{%
\partial z}}A_{z}(x,y,a,t)=0\,\,.
\end{equation}
The vector potential operator $\mathbf{\hat{A}}(\mathbf{r},t)$ that
satisfies the wave equation, the Coulmb gauge and the boundary conditions
can be written as
\begin{eqnarray}
\mathbf{\hat{A}}(\mathbf{r},t) &=&{\frac{1}{\pi }}\left( {\frac{\pi }{a}}%
\right) ^{\frac{1}{2}}\sum_{n=0}^{\;\;\infty\;\;\prime\prime}\int \,{\frac{d^{2}\mathbf{%
\kappa }}{\sqrt{\omega }}}\left\{ \hat{a}^{(1)}(\mathbf{\kappa },n)\mathbf{%
\hat{\kappa}}\times \hat{\mathbf{z}}\sin \left( {\frac{n\pi z}{a}}\right)
\right.  \notag \\
&+&\left. \hat{a}^{(2)}(\mathbf{\kappa },n)\left[ \mathbf{\hat{\kappa}}{%
\frac{in}{\omega a}}\sin \left( {\frac{n\pi z}{a}}\right) -\mathbf{\hat{z}}{%
\frac{\kappa }{\omega }}\cos \left( {\frac{n\pi z}{a}}\right) \right]
\right\} e^{i(\mathbf{\kappa }\cdot \mathbf{\rho}-\omega t)}\;+h.c,
\end{eqnarray}
where $\mathbf{\kappa }=(k_{x},k_{y})$ and $\mathbf{\rho }$ is the position
arrow on the $\mathcal{XY}$ plane. The normal frquencies are given by 
\begin{equation}
\omega =\omega (\mathbf{\kappa },n)=\sqrt{\mathbf{\kappa }^{2}+n^{2}{\frac{%
\pi ^{2}}{a^{2}}}},
\end{equation}
with $k_{x},k_{y}\in \mathbb{R}$ e $n\in \mathbb{N-}1$. The symbol $\sum^{"}$
indicates that the term corresponding to $n=0$ for normaliztion reasons
must be multiplied by $1/2$ . The Fourier coefficients $\hat{a}^{(\lambda
)}(\mathbf{\kappa },n)$ where $\lambda =1,2$ is \ the polarization index,
are operators \ in the photon ocupation number space and satisfy 
\begin{equation}
\left[ \hat{a}^{(\lambda )}(\mathbf{\kappa },n),\hat{a}^{\dagger (\lambda
^{\prime })}(\mathbf{\kappa }^{\prime },n^{\prime })\right] =\delta
_{\lambda \lambda ^{\prime }}\delta _{nn^{\prime }}\delta \left( \mathbf{%
\kappa }-\mathbf{\kappa }^{\prime }\right) \,.
\end{equation}
It is convenient to write the vector potential in the general form 
\begin{equation}
\mathbf{\hat{A}}(\mathbf{r},t)=\sum_{n=0}^{\;\;\infty\;\;
\prime\prime}\int \,{d^{2}}\mathbf{%
\kappa }\sum_{\lambda =1}^{2}\hat{a}^{(\lambda )}(\mathbf{\kappa },n)\mathbf{%
A}_{\mathbf{\kappa }n}^{(\lambda )}(\mathbf{r})e^{-i\omega (\mathbf{\kappa }%
,n)t}+h.c\,,
\end{equation}
where $\mathbf{A}_{\mathbf{\kappa }n}^{(\lambda )}(\mathbf{r})$ are the
modal functions. The modal functions for each polarization state must obey
Helmholtz equation and the boundary conditions given above. In our case the
modal functions are
\begin{equation}
\mathbf{A}_{\mathbf{\kappa \,}n}^{(1)}(\mathbf{r})={\frac{1}{\pi }}\left( {%
\frac{\pi }{a}}\right) ^{\frac{1}{2}}{\frac{1}{\sqrt{\omega }}}\sin \left( {%
\frac{n\pi z}{a}}\right) e^{-i\mathbf{\kappa }\cdot \mathbf{\rho }}\,\mathbf{%
\hat{\kappa}}\times \mathbf{\hat{z}}\,,  \label{AUM}
\end{equation}
and
\begin{equation}
\mathbf{A}_{\mathbf{\kappa \,}n}^{(2)}(\mathbf{r})={\frac{1}{\pi }}\left( {%
\frac{\pi }{a}}\right) ^{\frac{1}{2}}{\frac{1}{\sqrt{\omega }}}\left[ 
\mathbf{\hat{\kappa}}{\frac{in\pi }{a\omega }}\sin \left( {\frac{n\pi z}{a}}%
\right) -\mathbf{\hat{z}}{\frac{\kappa }{\omega }}\cos \left( {\frac{n\pi z}{%
a}}\right) \right] e^{-i\mathbf{\kappa }\cdot \mathbf{\rho }}\,.
\label{ADOIS}
\end{equation}
The next step is to evaluate the electric field operator $\mathbf{\hat{E}}(%
\mathbf{r},t)$. Recalling that $\hat{a}^{(\lambda )}(\mathbf{\kappa }%
,n)|0\rangle =0,$ we first write the correlators $<\hat{E}_{i}(\mathbf{r},t)%
\hat{E}_{j}(\mathbf{r},t)\rangle _{0}$ in the general form 
\begin{equation}
<\hat{E}_{i}(\mathbf{r},t)\hat{E}_{j}(\mathbf{r},t)\rangle _{0}=\sum_{\alpha
}E_{i\alpha }(\mathbf{r})E_{j\alpha }^{\ast }(\mathbf{r})\,,  \label{E0}
\end{equation}
where we have introduced the modal functions $E_{i\alpha }(\mathbf{r})$ for
the electric field. In our case (\ref{AUM}) and (\ref{ADOIS}) yield 
\begin{equation}
\mathbf{E}_{i\,\mathbf{\kappa }\,n}^{(1)}(\mathbf{r})={\frac{i}{\pi }}\left( 
{\frac{\omega \left( \mathbf{\kappa ,}n\right) \pi }{a}}\right) ^{\frac{1}{2}%
}\,\sin \left( {\frac{n\pi z}{a}}\right) e^{-i\mathbf{\kappa }\cdot \mathbf{%
\rho }}\,(\mathbf{\hat{\kappa}}\times \mathbf{\hat{z}})_{i}\,,  \label{E1}
\end{equation}
and, 
\begin{equation}
\mathbf{E}_{i\,\mathbf{\kappa \,}n}^{(2)}(\mathbf{r})={\frac{i}{\pi }}\left( 
{\frac{\omega \left( \mathbf{\kappa ,}n\right) \pi }{a}}\right) ^{\frac{1}{2}%
}\,\left[ \mathbf{\kappa }_{i}{\frac{in\pi }{a\,\omega \left( \mathbf{\kappa
,}n\right) }}\sin \left( \frac{n\pi z}{a}\right) -\mathbf{\hat{z}}_{i}{\frac{%
\kappa }{\omega \left( \mathbf{\kappa ,}n\right) }}\cos \left( \frac{n\pi z}{%
a}\right) \right] e^{-i\mathbf{\kappa }\cdot \mathbf{\rho }}\,,  \label{E2}
\end{equation}
respectively. Taking (\ref{E1}) and (\ref{E2}) into (\ref{E0}), we write $%
\mathbf{\hat{\kappa}}_{i}=\cos \phi \,\delta _{ix}+\sin \phi \,\delta _{iy}$%
, $\mathbf{\hat{z}}_{i}=\delta _{iz}$ e $(\mathbf{\hat{z}}\times \mathbf{%
\hat{\kappa}})_{i}=\sin \phi \,\delta _{ix}-\cos \phi \,\delta _{iy}$, where 
$\phi $ is the azimuthal angle on the $\mathcal{XY}$ plane and we have
performed all angular integrals. In this way we end up with 
\begin{eqnarray}
&&\langle {\hat{E}}_{i}(\mathbf{r},t){\hat{E}}_{j}(\mathbf{r},t)\rangle
_{0}\;\;=\;\;\left( {\frac{2}{\pi }}\right) \left( {\frac{\pi }{a}}\right) {%
\frac{\delta _{ij}^{\Vert }}{2}}\sum_{n=0}^{\;\;\infty\;\;\prime\prime}\sin ^{2}\left( \frac{%
n\pi z}{a}\right) \int_{0}^{\infty }d\kappa \,\kappa \,\omega (\mathbf{%
\kappa },n)  \notag \\
&+&\left( {\frac{2}{\pi }}\right) \left( {\frac{\pi }{a}}\right) \left( {%
\frac{\pi }{a}}\right) ^{2}{\frac{\delta _{ij}^{\Vert }}{2}}%
\sum_{n=0}^{\;\;\infty\;\;\prime\prime}n^{2}\sin ^{2}\left( \frac{n\pi z}{a}\right)
\int_{0}^{\infty }d\kappa \,\kappa \,\omega ^{-1}(\mathbf{\kappa },n)  \notag
\\
&+&\left( {\frac{2}{\pi }}\right) \left( {\frac{\pi }{a}}\right) \delta
_{ij}^{\bot }\sum_{n=0}^{\;\;\infty\;\;\prime\prime}\cos ^{2}\left( \frac{n\pi z}{a}\right)
\int_{0}^{\infty }d\kappa \,\kappa ^{3}\,\omega ^{-1}(\mathbf{\kappa },n)\,,
\label{VEV}
\end{eqnarray}
where $\delta _{ij}^{\Vert }:=\delta _{ix}\delta _{jx}+\delta _{iy}\delta
_{jy}$ e $\delta _{ij}^{\bot }:=\delta _{iz}\delta _{jz}$. Equation (\ref
{VEV}) is a formal expression for the correlator $\langle E_{i}(\mathbf{%
r},t)E_{j}(\mathbf{r},t)\rangle _{0}$, since it is mathematically
ill-defined unless a regularization recipe is prescribed. We will regularize
the integrals in (\ref{VEV}) with the help of analytical continuation
methods.

Consider, for instance, the first integral on the r.h.s. of (\ref{VEV}) and
let us rewrite it as follows
\begin{equation*}
\int_{0}^{\infty }d\kappa \kappa \left( \kappa ^{2}+{\frac{n^{2}\pi ^{2}}{%
a^{2}}}\right) ^{1/2}\rightarrow \int_{0}^{\infty }d\kappa \kappa \left(
\kappa ^{2}+{\frac{n^{2}\pi ^{2}}{a^{2}}}\right) ^{1/2-s}.\;
\end{equation*}
The first term in (\ref{VEV}) can be rewritten as
\begin{equation}
T_{1}=\left( {\frac{2}{\pi }}\right) \left( {\frac{\pi }{a}}\right) {\frac{%
\delta _{ij}^{\Vert }}{2}}\sum_{n=0}^{\infty "}\sin ^{2}\left( \frac{n\pi z}{%
a}\right) \int_{0}^{\infty }d\kappa \,\kappa \,\left( \mathbf{\kappa }^{2}+%
\frac{n^{2}\pi ^{2}}{a^{2}}\right) ^{-\frac{1}{2}-s},  \label{T1}
\end{equation}
the second as
\begin{equation}
T_{2}=\left( {\frac{2}{\pi }}\right) \left( {\frac{\pi }{a}}\right) \left( {%
\frac{\pi }{a}}\right) ^{2}{\frac{\delta _{ij}^{\Vert }}{2}}%
\sum_{n=0}^{\infty "}n^{2}\sin ^{2}\left( \frac{n\pi z}{a}\right)
\int_{0}^{\infty }d\kappa \,\kappa \left( \mathbf{\kappa }^{2}+\frac{%
n^{2}\pi ^{2}}{a^{2}}\right) ^{-\frac{1}{2}-s},  \label{T2}
\end{equation}
and the third one as
\begin{equation}
T_{3}=\left( {\frac{2}{\pi }}\right) \left( {\frac{\pi }{a}}\right) \delta
_{ij}^{\bot }\sum_{n=0}^{\infty "}\cos ^{2}\left( \frac{n\pi z}{a}\right)
\int_{0}^{\infty }d\kappa \,\kappa ^{3}\,\left( \mathbf{\kappa }^{2}+\frac{%
n^{2}\pi ^{2}}{a^{2}}\right) ^{-\frac{1}{2}-s}.  \label{T3}
\end{equation}
Let us assume that $\Re \,s$ is large enough to give precise mathematical
meaning to these integrals. After evaluating them and make use of the
analytical continuation of the results we will take the limit $s\rightarrow 0$.
Let us see, for instance, what happens with $T_{1}$. Making use of the
following representation of Euler beta function \cite{Grad}
\begin{equation}
\int_{0}^{\infty }dx\,x^{\mu -1}\left( x^{2}+c^{2}\right) ^{\nu -1}={\frac{B%
}{2}}\left( {\frac{\mu }{2}},1-\nu -{\frac{\mu }{2}}\right) c^{\mu +2\nu
-2}\,,
\end{equation}
where $B(x,y)=\Gamma (x)\Gamma (y)/\Gamma (x+y)$, and that holds for $\Re
\,\left( \nu +{\frac{\mu }{2}}\right) <1$ and $\Re \,\mu >0$, we obtain 
\begin{equation}
\int_{0}^{\infty }d\kappa \kappa \left( \kappa ^{2}+{\frac{n^{2}\pi ^{2}}{%
a^{2}}}\right) ^{1/2-s}={\frac{1}{2}}\left( {\frac{n\pi }{a}}\right)
^{3-2s}\;{\frac{\Gamma (s-3/2)}{\Gamma (s-1/2)}}={\frac{1}{(2s-3)}}\left( {%
\frac{n\pi }{a}}\right) ^{3-2s}\;.
\end{equation}
\noindent Taking this result into $T_{1}$ we obtain
\begin{equation}
T_{1}=\left( {\frac{1}{2s-3}}\right) \left( {\frac{\pi }{a}}\right) ^{3-2s}{%
\frac{\delta _{ij}^{\Vert }}{2a}}\left[ \zeta _{R}(2s-3)-\sum_{n=0}^{\infty
}n^{3-2s}\cos \left( {\frac{2n\pi z}{a}}\right) \right] \;,
\end{equation}
where $\zeta _{R}(z)$ is the well-known Riemann zeta function. Taking the
limit $s\rightarrow 0$, we have
\begin{equation}
T_{1}=-{\frac{1}{3\pi }}\left( {\frac{\pi }{a}}\right) ^{4}\frac{\delta
_{ij}^{\Vert }}{2}\left[ {\frac{1}{120}}+\frac{1}{8}\sum_{n=1}^{\infty }%
\frac{d^{3}}{d{\xi }^{{3}}}\sin \left( 2n\xi \right) \right] \;,
\label{T1prima}
\end{equation}
where we made use of the fact that $\zeta _{R}(-3)=1/120$, defined $\xi :=\pi z/a$, and wrote
\begin{equation}
n^{3}\cos \left( 2n\xi \right) =-\frac{1}{8}\times {\frac{d^{3}}{d\xi ^{3}}%
\sin }\left( 2n\xi \right).
\end{equation}
The sum on the R.H.S. of (\ref{T1prima}) can be regularized in many ways. A
quick though non-rigorous way is to write
\begin{equation}
T_{1}=-{\frac{1}{3\pi }}\left( {\frac{\pi }{a}}\right) ^{4}\frac{\delta
_{ij}^{\Vert }}{2}\left[ {\frac{1}{120}}+\frac{1}{8}\frac{d^{3}}{d{\xi }^{{3}%
}}\sum_{n=1}^{\infty }\sin \left( 2n\xi \right) \right] ,\;
\label{T1primaprima}
\end{equation}
and express the summand in terms of exponential functions of imaginary
argument thereby transforming each one of the sums into
the euclidean space. In this way we obtain
\begin{eqnarray}
\sum_{n=1}^{\infty }\sin \left( 2n\xi \right)  &=&\frac{1}{2i}\left(
\sum_{n=1}^{\infty }\exp \left( i2n\xi \right) -\sum_{n=1}^{\infty }\exp
\left( -i2n\xi \right) \right)   \notag \\
&=&\frac{1}{2i}\left( \sum_{n=1}^{\infty }\exp \left( 2n\xi _{E}\right)
-\sum_{n=1}^{\infty }\exp \left( -2n\xi _{E}^{\prime }\right) \right) 
\end{eqnarray}
where we have made the substitution $i\xi \to -\xi _{E}$ in the first
sum and $i\xi \to \xi _{E}^{\prime }$ in the second. Each one of the
sums above can be easily performed and the result is
\begin{equation}
\sum_{n=1}^{\infty }\sin \left( 2n\xi \right) =\frac{1}{2}\cot \left( \xi
\right) 
\end{equation}
It follows that 
\begin{equation}
T_{1}=-{\frac{1}{3\pi }}\left( {\frac{\pi }{a}}\right) ^{4}\frac{\delta
_{ij}^{\Vert }}{2}\left[ {\frac{1}{120}}+\frac{1}{8}\frac{d^{3}}{d{\xi }^{{3}%
}}\frac{1}{2}\cot \left( \xi \right) \right].
\end{equation}
Treating the two other terms in (\ref{VEV}) in a similar manner we obtain
\begin{equation}
T_{2}=-{\frac{1}{\pi }}\left( {\frac{\pi }{a}}\right) ^{4}\frac{\delta
_{ij}^{\Vert }}{2}\left[ {\frac{1}{120}}+\frac{1}{8}\frac{d^{3}}{d{\xi }^{{3}%
}}\frac{1}{2}\cot \left( \xi \right) \right],
\end{equation}
and
\begin{equation}
T_{3}={\frac{4}{3\pi }}\left( {\frac{\pi }{a}}\right) ^{4}\frac{\delta
_{ij}^{\bot }}{2}\left[ {\frac{1}{120}}-\frac{1}{8}\frac{d^{3}}{d{\xi }^{{3}}%
}\frac{1}{2}\cot \left( \xi \right) \right].
\end{equation}
Notice that some care must be taken when we aply this procedure to the
third term. This is so because the term corresponding to $n=0$ in $T_{3}$ is
not zero. In fact its contribution is: 
\begin{equation}
T_{3}(n=0)=\left( {\frac{2}{\pi }}\right) \left( {\frac{\pi }{a}}\right)
\delta _{ij}^{\bot }\int_{0}^{\infty }d\kappa \,\kappa \,^{-2-2s},
\end{equation}
which diverges when the regularization is removed. How ever this term is
non-physical and can be safely ignored. Finally, collecting all partial
results we have
\begin{eqnarray}
\langle E_{i}(\mathbf{r},t)E_{j}(\mathbf{r},t)\rangle _{0}
&=&T_{1}+T_{2}+T_{3}  \notag \\
&=&\left( {\frac{\pi }{a}}\right) ^{4}{\frac{2}{3\pi }}\left[ \left( -\delta
^{\Vert }+\delta ^{\bot }\right) _{ij}\;{\frac{1}{120}}+\delta _{ij}F(\xi )%
\right] \,.  \label{ECORRCASIMIR}
\end{eqnarray}
The function $\,F\left( \xi \right) $ is defined by 
\begin{equation}
F\left( \xi \right) :=-\frac{1}{8}\frac{d^{3}\,}{d\xi ^{3}}\frac{1}{2}\cot
\left( \xi \right),
\end{equation}
and its expansion about $\xi =0$ is given by
\begin{equation}\label{Fapp}
F\left( \xi \right) \approx \frac{3}{8}\xi^{-4}+\frac{1}{120}+O\left(\xi^2\right).
\end{equation}
Near $\xi =\pi $ (which corresponds to $z=a$) we make the
replacement $\xi \rightarrow \xi -\pi $. Notice that due to the behavior of $%
F\left( \xi \right) $ near $\xi =0,a$, strong divergences predominate in the
behavior of the correlators near the plates.

By applying the exactly the same procedure we obtain the magnetic field
correlators
\begin{equation}
\langle B_{i}(\mathbf{r},t)B_{j}(\mathbf{r},t)\rangle _{0}=\left( {\frac{\pi 
}{a}}\right) ^{4}{\frac{2}{3\pi }}\left[ \left( -\delta ^{\Vert }+\delta
^{\bot }\right) _{ij}\;{\frac{1}{120}}-\delta _{ij}F(\xi )\right] \,.
\label{BCORRCASIMIR}
\end{equation}
A direct evaluation also shows that the correlators $<E_{i}(\mathbf{r},t)B_{j}(\mathbf{r},t)\rangle _{0}$ are zero.

\section{Correlators for Boyer's setup}

\bigskip The other setup we are interested in is that one in which a
perfectly conducting plate is placed at $z=0$ and perfectly permeable plate
is placed at $z=a$. This setup was analyzed for the first time by Boyer in
the contetxt of stochastic electrodynamic \cite{Boyer74}\ and it is the
simplest case of a repulsive Casimir effect that can be found in the
literature. The boundary conditions now are: \emph{(a)} the tangential
components $E_{x}$ and $E_{y}$ of the electric field as well as the normal
component $B_{z}$ of the magnetic field must vanish on the surface of the
plate at $z=0$; \emph{(b)} the tangential components of $B_{x}$ e $B_{y}$ of
the magnetic field as well as normal conponent $E_{z}$\ of the electric
field must vanish on the surface of the plate at $z=a$. These boundary
conditions translated in terms of the components of the vector potential
read
\begin{equation}
A_{x}(x,y,0,t)=0\,;\;\;\;\;A_{y}(x,y,0,t)=0\,;\;\;\;\;{\frac{\partial }{%
\partial z}}A_{z}(x,y,0,t)=0\,,
\end{equation}
at $\ z=0$, and at $z=a$
\begin{equation}
{\frac{\partial }{\partial x}}A_{x}(x,y,a,t)=0\,;\;\;\;\;{\frac{\partial }{%
\partial y}}A_{y}(x,y,a,t)=0\,;\;\;\;\;A_{z}(x,y,a,t)=0\,.
\end{equation}
The appropriate vector potential operator $\mathbf{\hat{A}}(\mathbf{r},t)$
is given by \cite{JPA99}
\begin{eqnarray}
\mathbf{\hat{A}}(\mathbf{r},t) &=&{\frac{1}{\pi }}\left( {\frac{\pi }{a}}%
\right) ^{\frac{1}{2}}\sum_{n=0}^{\infty }\int \,{\frac{d^{2}\mathbf{\kappa }%
}{\sqrt{\omega }}}\left\{ \hat{a}^{(1)}(\mathbf{\kappa },n)\mathbf{\hat{%
\kappa}}\times \hat{\mathbf{z}}\sin \left[ \left( n+{\frac{1}{2}}\right) {%
\frac{\pi z}{a}}\right] \right.  \notag \\
&+&\left. \hat{a}^{(2)}(\mathbf{\kappa },n)\left[ \mathbf{\hat{\kappa}}{%
\frac{i(n+{\frac{1}{2}})}{\omega a}}\sin \left[ \left( n+{\frac{1}{2}}%
\right) {\frac{\pi z}{a}}\right] -\mathbf{\hat{z}}{\frac{\kappa }{\omega }}%
\cos \left[ \left( n+{\frac{1}{2}}\right) {\frac{\pi z}{a}}\right] \right]
\right\}  \notag \\
&&\times e^{i(\mathbf{\kappa }\cdot \mathbf{\rho }-\omega t)}\;+\text{ }h.c.,
\end{eqnarray}
where as before $\mathbf{\kappa }=(k_{x},k_{y})$ and $\mathbf{\rho }$ is the
position vector on the $\mathcal{XY}$ plane. The normal frequencies are
given 
\begin{equation}
\omega (\mathbf{\kappa },n)=\sqrt{\mathbf{\kappa }^{2}+\left( n+{\frac{1}{2}}%
\right) ^{2}{\frac{\pi ^{2}}{a^{2}}}}\,,
\end{equation}
with $k_{x},k_{y}\in \mathbb{R}$ and $n\in \mathbb{N-}1$. Notice that
contrary to the case of two conducting plates normalization does not require
the we multiply the term corresponding to $n=0$ by 
$1/2$. The electric and magnetic field correlators for Boyer's setup can be
evaluated with the same technique employed before \cite{JPA99}. In fact, it
is not hard to convince ourselves that it is sufficient to perform the
substitution $n\to n+1/2$ \ and follow the same steps as before to
obtain
\begin{equation}
\langle E_{i}(\mathbf{r},t)E_{j}(\mathbf{r},t)\rangle _{0}=T_{1}+T_{2}+T_{3},
\end{equation}
where, for example
\begin{equation}
T_{1}=\frac{\Gamma \left( s-\frac{3}{2}\right) }{\Gamma \left( s-\frac{1}{2}%
\right) }\left( {\frac{\pi }{a}}\right) ^{3-2s}{\frac{\delta _{ij}^{\Vert }}{%
2a}}\left\{ \sum_{n=0}^{\infty }\left( n+\frac{1}{2}\right) ^{3-2s}\frac{1}{2%
}\left[ 1-\cos \left( 2\left( n+1/2\right) \frac{\pi z}{a}\right) \right]
\right\} ,\;
\end{equation}
The terms $T_{2}$ e $T_{1}$ show a similar structure. The main difference
with respect to the two conducting plate case is that now we have to deal
with the Hurwitz zeta function $\zeta _{H}(z,q)$ which has a series
representation given by \cite{Grad}
\begin{equation}
\zeta _{H}(z,q)=\sum_{n=0}^{\infty }\frac{1}{\left( n+q\right) ^{z}},
\end{equation}
with $\Re \,z>1$, and $q\neq 0,-1,-2,....$. In our case we must set
\begin{equation}
\zeta _{H}(2s-3,\frac{1}{2})=\sum_{n=0}^{\infty }\left( n+\frac{1}{2}\right)
^{3-2s}.
\end{equation}
It follows that in the limit $s\rightarrow 0$ we have
\begin{equation}
T_{1}=-\left( {\frac{\pi }{a}}\right) ^{4}\frac{1}{3\pi }{\frac{\delta
_{ij}^{\Vert }}{2}}\left\{ \zeta _{H}\left( -3,\frac{1}{2}\right)
+\sum_{n=0}^{\infty }\left( n+\frac{1}{2}\right) ^{3}\cos \left( 2\left( n+%
\frac{1}{2}\right) \frac{\pi z}{a}\right) \right\} .
\end{equation}
In the same way we obtain for $T_{2}$ e $T_{1}$ the results
\begin{equation}
T_{2}=-\left( {\frac{\pi }{a}}\right) ^{4}\frac{1}{\pi }{\frac{\delta
_{ij}^{\Vert }}{2}}\left\{ \zeta _{H}\left( -3,\frac{1}{2}\right)
-\sum_{n=0}^{\infty }\left( n+\frac{1}{2}\right) ^{3}\cos \left( 2\left( n+%
\frac{1}{2}\right) \frac{\pi z}{a}\right) \right\} .
\end{equation}
and
\begin{equation}
T_{3}=\left( {\frac{\pi }{a}}\right) ^{4}\frac{2}{3\pi }{\frac{\delta
_{ij}^{\bot }}{2}}\left\{ \zeta _{H}\left( -3,\frac{1}{2}\right)
+\sum_{n=0}^{\infty }\left( n+\frac{1}{2}\right) ^{3}\cos \left( 2\left( n+%
\frac{1}{2}\right) \frac{\pi z}{a}\right) \right\} .
\end{equation}
Notice that this time we do not have the divergent contribution corresponding
to $n=0$ as in the case of the conducting plates. Adding the three terms we
have 
\begin{eqnarray}
\left\langle \hat{E}_{i}\left( \mathbf{r},t\right) \hat{E}_{j}\left( \mathbf{%
r},t\right) \right\rangle _{0} &=&\left( {\frac{\pi }{a}}\right) ^{4}\frac{2%
}{3\pi }\left\{ (-\delta _{ij}^{\Vert }+\delta _{ij}^{\bot })\zeta
_{H}\left( -3,\frac{1}{2}\right) \right.   \notag \\
&&\left. +\sum_{n=0}^{\infty }\left( n+\frac{1}{2}\right) ^{3}\cos \left(
2\left( n+\frac{1}{2}\right) \frac{\pi z}{a}\right) \right\} .
\end{eqnarray}
The numerical value $\zeta _{H}\left( -3,\frac{1}{2}\right) $ can be
obtained from \cite{Grad} 
\begin{equation}
\zeta _{H}\left( -n,q\right) =-\frac{B_{n+1}(q)}{n+1},
\end{equation}
where $n\in \mathbf{N}$ and $B_{n+1}(q)$ is a Bernoulli polynomial defined
by \cite{Grad} 
\begin{equation}
B_{n}(x)=\sum_{p=0}^{n}\frac{n!}{p!(n-p)!}B_{p}x^{n-k},
\end{equation}
where $B_{p}$ is a Bernoulli number. The relevant polynomial here is: 
\begin{equation}
B_{4}(x)=x^{4}-2x^{3}+x^{2}-\frac{1}{30}.
\end{equation}
With $B_{4}(1/2)=(7/8)\times (1/30)$, it follows that $\zeta _{H}\left( -3,%
\frac{1}{2}\right) =-(7/8)(1/120)$. The sum can be regularized with the same
technique employed before. In fact, we can define the function $G(\xi )$ by 
\begin{eqnarray}
G(\xi ) &:&=\sum_{n=0}^{\infty }\left( n+\frac{1}{2}\right) ^{3}\cos \left[
2\left( n+\frac{1}{2}\right) \xi \right]   \notag \\
&=&\frac{1}{8}\sum_{n=0}^{\infty }\left( 2n+1\right) ^{3}\cos \left[ \left(
2n+1\right) \xi \right] ,
\end{eqnarray}
where as before $\xi :=z\pi /a$. We can write 
\begin{equation}
\left( 2n+1\right) ^{3}\cos \left[ \left( 2n+1\right) \xi \right] =-\frac{%
d^{3}}{d\xi ^{3}}\sin \left[ \left( 2n+1\right) \xi \right] 
\end{equation}
and formally we have 
\begin{equation}
G(\xi )=-\frac{1}{8}\frac{d^{3}}{d\xi ^{3}}\sum_{n=0}^{\infty }\sin \left[
\left( 2n+1\right) \xi \right] .
\end{equation}
Writing $\sin \left[ \left( 2n+1\right) \xi \right] $ in terms of
exponentials of imaginary argument and passing to the euclidean space we
obtain after some simple manipulations
\begin{equation}
\,G\left( \xi \right) =-\frac{1}{8}\frac{d^{3}\,}{d\xi ^{3}}\frac{1}{2\sin
\left( \xi \right) }.
\end{equation}
Collecting all partial results we finally obtain for $\left\langle \hat{E}%
_{i}\left( \mathbf{r},t\right) \hat{E}_{j}\left( \mathbf{r},t\right)
\right\rangle _{0}$ the result
\begin{equation}
\left\langle \hat{E}_{i}\left( \mathbf{r},t\right) \hat{E}_{j}\left( \mathbf{%
r},t\right) \right\rangle _{0}=\left( \frac{\pi }{a}\right) ^{4}\frac{2}{%
3\pi }\left[ \left( -\frac{7}{8}\right) \frac{\left( -\delta ^{\Vert
}+\delta ^{\perp }\right) _{ij}}{120}+\delta _{ij}\,G\left( \xi \right) %
\right].   \label{EcorrBoyer}
\end{equation}
Proceeding in the same way in the evaluation of  $\left\langle \hat{B}%
_{i}\left( \mathbf{r},t\right) \hat{B}_{j}\left( \mathbf{r},t\right)
\right\rangle _{0}$ we obtain
\begin{equation}
\left\langle \hat{B}_{i}\left( \mathbf{r},t\right) \hat{B}_{j}\left( \mathbf{%
r},t\right) \right\rangle _{0}=\left( \frac{\pi }{a}\right) ^{4}\frac{2}{%
3\pi }\left[ \left( -\frac{7}{8}\right) \frac{\left( -\delta ^{\Vert
}+\delta ^{\perp }\right) _{ij}}{120}-\delta _{ij}\,G\left( \xi \right) %
\right].   \label{BcorrBoyer}
\end{equation}
Observe that near $\xi =0$ the function $G\left( \xi \right) $ behaves as
\begin{equation}
G\left( \xi \right) =\frac{3}{8}\xi ^{-4}-\frac{7}{8}\,\frac{1}{120}+O\left(
\xi ^{2}\right) ,  \label{Gapp1}
\end{equation}
But near $\xi =\pi $ its behavior is slightly different
\begin{equation}
G\left( \xi \right) =-\frac{3}{8}\left( \xi -\pi \right) ^{-4}+\frac{7}{8}\,%
\frac{1}{120}+O\left[ \left( \xi -\pi \right) ^{2}\right] .  \label{Gapp2}
\end{equation}
Again, a direct calculation shows that $\left\langle \hat{E}_{i}\left( 
\mathbf{r},t\right) \hat{B}_{j}\left( \mathbf{r},t\right) \right\rangle
_{0}=0$ for this case.

As before the divergent behavior of the correlators near the plates we are
interested in is an effect of the distortions of the electromagnetic
oscillations with respect to a situation where the plates are not present.
This fact has received the attention of several authors, see for example \cite
{Deutsch79,Lü&Ravndal85}.

\section{The Casimir effect for conducting plates}

In order to apply the above results to Casimir's original setup we consider
for convenience three parallel perfectly conducting plates perpendicular to
the $\mathcal{OZ}$ axis at $z=0$, $z=a$ and $z=\ell $.
\begin{figure}\label{3plate}
\begin{center}
\psfig{figure=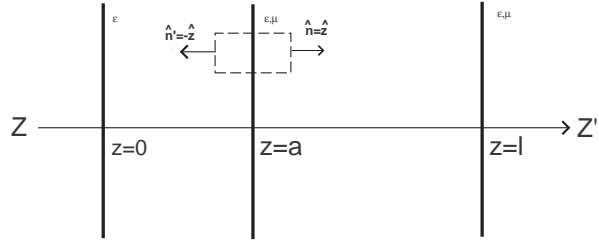,width=10cm,angle=0}
\caption{Three-plate setup for the obtention of the Casimir force per unit area. The plate at $z=\ell$ is auxiliary.}
\end{center}
\end{figure}
The quantum version of Maxwell tensor is 
\begin{equation}
4\pi \left\langle \hat{T}_{ij}\right\rangle _{0}=\left\langle \hat{E}_{i}\,%
\hat{E}_{j}\right\rangle _{0}-\frac{1}{2}\,\delta _{ij}\left\langle \mathbf{%
\hat{E}}^{2}\right\rangle _{0}+\left\langle \hat{B}_{i}\,\hat{B}%
_{j}\right\rangle _{0}-\frac{1}{2}\,\delta _{ij}\left\langle \mathbf{\hat{B}}%
^{2}\right\rangle _{0}
\end{equation}
Making use of the correlator given by (\ref{ECORRCASIMIR}) we obtain the
following partial results 
\begin{equation}
\left\langle \hat{E}_{x}^{2}(z,t)\right\rangle _{0}=\left\langle \hat{E}%
_{y}^{2}(z,t\right\rangle _{0}=\left( \frac{\pi }{a}\right) ^{4}\frac{2}{%
3\pi }\left[ -\frac{1}{120}+F\left( \xi \right) \right] ,
\end{equation}
\begin{equation}
\left\langle \hat{E}_{z}^{2}(z,t\right\rangle _{0}=\left( \frac{\pi }{a}%
\right) ^{4}\frac{2}{3\pi }\left[ \frac{1}{120}+F\left( \xi \right) \right] ,
\end{equation}
also $\left\langle \hat{E}_{x}(z,t)\hat{E}_{y}(z,t)\right\rangle
_{0}=\left\langle \hat{E}_{x}(z,t)\hat{E}_{z}(z,t)\right\rangle
_{0}=\left\langle \hat{E}_{y}(z,t)\hat{E}_{z}(z,t)\right\rangle _{0}=0$. In
the same way, making use of (\ref{BCORRCASIMIR}) we obtain 
\begin{equation}
\left\langle \hat{B}_{x}^{2}(z,t)\right\rangle _{0}=\left\langle \hat{B}%
_{y}^{2}(z,t\right\rangle _{0}=\left( \frac{\pi }{a}\right) ^{4}\frac{2}{%
3\pi }\left[ -\frac{1}{120}-F\left( \xi \right) \right] ,
\end{equation}
\begin{equation}
\left\langle \hat{B}_{z}^{2}(z,t\right\rangle _{0}=\left( \frac{\pi }{a}%
\right) ^{4}\frac{2}{3\pi }\left[ \frac{1}{120}-F\left( \xi \right) \right] ,
\end{equation}
and also $\left\langle \hat{B}_{x}(z,t)\hat{B}_{y}(z,t)\right\rangle
_{0}=\left\langle \hat{B}_{x}(z,t)\hat{B}_{z}(z,t)\right\rangle
_{0}=\left\langle \hat{B}_{y}(z,t)\hat{B}_{z}(z,t)\right\rangle _{0}=0$. The
components of the quantum version of Maxwell tensor can be easily evaluated.
For instance 
\begin{eqnarray}
8\pi \left\langle \hat{T}_{zz}(z,t)\right\rangle _{0} &=&\left\langle \hat{E}%
_{z}^{2}(z,t)\right\rangle _{0}-\left\langle \hat{E}_{x}^{2}(z,t)\right%
\rangle _{0}-\left\langle \hat{E}_{y}^{2}(z,t)\right\rangle _{0}  \notag \\
&&+\left\langle \hat{B}_{z}^{2}(z,t)\right\rangle _{0}-\left\langle \hat{B}%
_{x}^{2}(z,t)\right\rangle _{0}-\left\langle \hat{B}_{y}^{2}(z,t)\right%
\rangle _{0}.
\end{eqnarray}
performing the necessary substitutions we have 
\begin{equation}
\left\langle \hat{T}_{zz}\right\rangle _{0}=\frac{\pi ^{2}}{240a^{4}}.
\label{ZZpressure}
\end{equation}
In the same way 
\begin{eqnarray}
\left\langle \hat{T}_{xx}\right\rangle _{0} &=&\left\langle \hat{T}%
_{yy}\right\rangle _{0}=-\frac{1}{8\pi }\left( \left\langle \hat{E}%
_{z}^{2}(z,t)\right\rangle _{0}+\left\langle \hat{B}_{z}^{2}(z,t)\right%
\rangle _{0}\right)   \notag \\
&=&-\frac{\pi ^{2}}{720a^{4}}  \label{XXpressure}
\end{eqnarray}
Notice how conveniently the divergent parts near tha plates cancel out
yielding finite results.

Let us obtain now the Casimir force per unit area between the conducting
plates. Consider Figure (\ref{3plate}) and  the plate at  $z=a$. The
Casimir force per unit area on this plate is 
\begin{eqnarray}
\frac{F_{z}}{A} &=&-T_{zz}(z\rightarrow a^{L})+T_{zz}(z\rightarrow a^{R}) 
\notag \\
&=&-\frac{\pi ^{2}}{240a^{4}}+\frac{\pi ^{2}}{240\left( \ell -a\right) ^{4}},
\end{eqnarray}
where $z\rightarrow a^{L,R}$ means that $z$ tends to $a$ from the left/right.
Taking the limit $\ell \rightarrow \infty $ we obtain the expected result
for the Casimir force per unit area 
\begin{equation}
\frac{F_{z}}{A}=-\frac{\pi ^{2}}{240a^{4}}.
\end{equation}
The minus sign shows that the resulting pressure pushes towards the region
between the plates. If simultaneously we take the limits $\ell ,a\rightarrow
\infty $ keeping the distance $\ell -a$ constant. The pressure changes its
sign but it still pushes the plate at $z=a$ towards the one at $z=\ell $.

In order to calculate the renormalized symmetrical stress-energy tensor $%
\langle {}\,\hat{\Theta}^{\mu \nu }\left( z\right) \rangle _{0}^{ren}$%
\thinspace\ we evaluate the energy density $\rho \left( z\right) \equiv
\langle {}\,\hat{\Theta}^{00}\left( z\right) \,\rangle _{0}^{_{ren}}$ in the
region between the plates as well as the saptial components $\langle {}\hat{%
\Theta}^{xx}\left( z\right) \rangle _{0}^{ren}$, $\langle {}\hat{\Theta}%
^{yy}\left( z\right) \rangle _{0}^{ren}$, and $\langle {}\hat{\Theta}%
^{zz}\left( z\right) \rangle _{0}^{ren}$. The energy density is given by 
\begin{equation}
\rho \left( \mathbf{r},t\right) =\frac{1}{8\pi }\left( \left\langle \mathbf{%
\hat{E}}^{2}\left( \mathbf{r},t\right) \right\rangle _{0}+\left\langle 
\mathbf{\hat{B}}^{2}\left( \mathbf{r},t\right) \right\rangle _{0}\right)
\end{equation}
making use of the correlators given by (\ref{ECORRCASIMIR}) and (\ref
{BCORRCASIMIR}) we obtain 
\begin{equation}
\rho (a)=-\frac{\pi ^{2}}{720a^{4}}.
\end{equation}
This result is due to the fact that the divergent pieces in (\ref
{ECORRCASIMIR}) and (\ref{BCORRCASIMIR}) cancel out yielding a finite result
for the vacuum energy density. Recalling that $\Theta ^{ij}(z)=-T_{ij}(z)$,
see \cite{Jackson3rd}, with help of (\ref{MaxwellStressTensor}), (\ref
{ECORRCASIMIR}) and (\ref{BCORRCASIMIR}) the remanescent components of the
symmetrical stress-energy tensor are easily obtained. The final result is 
\begin{equation}
\langle {}\hat{\Theta}^{\mu \nu }(z)\rangle _{0}^{_{ren}}=\frac{\pi ^{2}}{%
720a^{4}}\mbox{diag}\,\left( -1,1,1,-3\right) ,  \label{B&Mac}
\end{equation}
which is perfect agreement with Brown and Maclay's results \cite
{Brown&Maclay69}. \ Notice also that $\langle {}\hat{\Theta}_{\mu }^{\mu
}(z)\rangle _{0}^{_{ren}}=g_{\mu \nu }\langle {}\hat{\Theta}^{\mu \nu
}(z)\rangle _{0}^{_{ren}}=0$, with $g_{\mu \nu }=\mbox{diag}(1,-1,-1,-1)$.

\section{The Casimir effect for one conducting plate and an infinitely
permeable one}

Let us consider now the setup proposed by Boyer \cite{Boyer74} which
consists of a perfectly conducting plate placed perpendicularly to the $%
\mathcal{OZ}$ axis at $z=0$ and another infinitely permeable one parallel to
the first placed at  $z=a$. The boundary conditions on the conducting
plate are as before $E_{x}=E_{y}=0$ and $B_{z}=0$, and for the infinitely
permeable plate: $B_{x}=B_{y}=0$ and $E_{z}=0$. Making use of the correlator
given by  (\ref{EcorrBoyer}) the following partial results:
\begin{equation}
\left\langle \hat{E}_{x}^{2}(z,t)\right\rangle _{0}=\left\langle \hat{E}%
_{y}^{2}(z,t\right\rangle _{0}=\left( \frac{\pi }{a}\right) ^{4}\frac{2}{%
3\pi }\left[ \frac{7}{8}\times \frac{1}{120}+G\left( \xi \right) \right] ,
\end{equation}
\begin{equation}
\left\langle \hat{E}_{z}^{2}(z,t\right\rangle _{0}=\left( \frac{\pi }{a}%
\right) ^{4}\frac{2}{3\pi }\left[ \left( -\frac{7}{8}\right) \times \frac{1}{%
120}+G\left( \xi \right) \right] ,
\end{equation}
and $\left\langle \hat{E}_{x}(z,t)\hat{E}_{y}(z,t)\right\rangle
_{0}=\left\langle \hat{E}_{x}(z,t)\hat{E}_{z}(z,t)\right\rangle
_{0}=\left\langle \hat{E}_{y}(z,t)\hat{E}_{z}(z,t)\right\rangle _{0}=0$. By
the same token making use of the correlator given by (\ref{BcorrBoyer}) we
obtain 
\begin{equation}
\left\langle \hat{B}_{x}^{2}(z,t)\right\rangle _{0}=\left\langle \hat{B}%
_{y}^{2}(z,t\right\rangle _{0}=\left( \frac{\pi }{a}\right) ^{4}\frac{2}{%
3\pi }\left[ \frac{7}{8}\times \frac{1}{120}-G\left( \xi \right) \right] ,
\end{equation}
\begin{equation}
\left\langle \hat{B}_{z}^{2}(z,t\right\rangle _{0}=\left( \frac{\pi }{a}%
\right) ^{4}\frac{2}{3\pi }\left[ \left( -\frac{7}{8}\right) \times \frac{1}{%
120}-G\left( \xi \right) \right],
\end{equation}
and also 
\begin{equation}
\left\langle \hat{B}_{x}(z,t)\hat{B}_{y}(z,t)\right\rangle _{0}=\left\langle 
\hat{B}_{x}(z,t)\hat{B}_{z}(z,t)\right\rangle _{0}=\left\langle \hat{B}%
_{y}(z,t)\hat{B}_{z}(z,t)\right\rangle _{0}=0
\end{equation}
Proceeding as in the case of the conducting plates we obtain the following
results for the componenets of the quantum version of Maxwell tensor 
\begin{equation}
\left\langle \hat{T}_{xx}\right\rangle _{0}=\left\langle \hat{T}%
_{yy}\right\rangle _{0}=\frac{7}{8}\times \frac{\pi ^{2}}{720a^{4}},
\end{equation}
and 
\begin{equation}
\left\langle \hat{T}_{zz}\right\rangle _{0}=\left( -\frac{7}{8}\right)
\times \frac{\pi ^{2}}{240a^{4}}.
\end{equation}
Notice that it is sufficient to multiply the results obtained for Casimir's
setup by the factor $(-7/8)$ in order to obtain the results corresponding to
Boyer's setup.

In order to obtain the Casimir force per unit area for this setup it is
convenient to place a third conducting plate at $z=\ell $. \ Then the
Casimir force per unit area on the plate at $z=a$ will be given by 
\begin{eqnarray}
\frac{F_{z}}{A} &=&-T_{zz}(z\rightarrow a^{L})+T_{zz}(z\rightarrow a^{R}) 
\notag \\
&=&-\left( -\frac{7}{8}\right) \times \frac{\pi ^{2}}{240a^{4}}+\left( -%
\frac{7}{8}\right) \times \frac{\pi ^{2}}{240\left( \ell -a\right) ^{4}}.
\end{eqnarray}
Taking the limit $\ell \rightarrow \infty $ we obtain a Casimir force which
pushes the plate at $z=a$ towards the region $z>a$ given by 
\begin{equation}
\frac{F_{z}}{A}=\frac{7}{8}\times \frac{\pi ^{2}}{240a^{4}}
\end{equation}
This is the result obtained by Boyer \cite{Boyer74} for this setup using
stochastic electrodynamic methods and it is one of the simplest example of a
repulsive Casimir force.

In order to evaluate the symmetrical stress-energy tensor we first evaluate
the Casimir energy density. Making use of (\ref{EcorrBoyer}) e (\ref
{BcorrBoyer}) we have 
\begin{equation}
\rho =\frac{7}{8}\times\frac{\pi^2}{240a^4}.
\end{equation}

As in the case of the conducting plates a simple calculation shows that the
stress energy tensor for Boyer's setup is given by 
\begin{equation}
\langle {}\hat{\Theta}^{\mu \nu }(z)\rangle _{0}^{_{ren}}=\frac{7}{8}%
\,\times \frac{\pi ^{2}}{720a^{4}}\mbox{diag}\,\left( 1,-1,-1,3\right) .
\end{equation}
As before $\langle {}\hat{\Theta}_{\mu }^{\mu }(z)\rangle
_{0}^{_{ren}}=g_{\mu \nu }\langle {}\hat{\Theta}^{\mu \nu }(z)\rangle
_{0}^{_{ren}}=0$.

\section{The interaction between an atom and two material surfaces}

In 1948, Casimir and Polder \cite{CasimirPolder48} taking into account a
suggestion made by experimentalists evaluated the interaction potential
between two eletrical polarizable molecules separated by a distance $r$
including the effects due to the finiteness of the speed of propagation of
the electromagnetic interaction, i.e.: of the retardment. Casimir and Polder
showed that the retardment causes the interaction potential to change from a 
$r^{-6}$ power law to a  $r^{-7}$ power law. In the same paper, Casimir and
Polder also analyzed the interaction between an atom and a conducting wall e
showed the interaction potetntial in this case varies according to a  $r^{-4}
$,  where now  $r$ is the distance between the atom and the wall. Here we
will show how it is possible to reobtain with the help of the correlators
given by (\ref{ECORRCASIMIR}) and (\ref{BCORRCASIMIR}),the piece of Casimir
and Polder's result for the atom-wall interaction that depends on the
distortion of the vaccum oscillations of the electromagnetic field caused by
the presence of the wall.

From a classical point of view the induced eletrical polarization density $%
\mathbf{P}$ can be thought of as a function of the electric amd magnetic
fields $\mathbf{E}$ and $\mathbf{B}$. In many cases only the dependence on
eletric field is relevant. It can be shown that under conditions for which
the effects of the retardment must be taken into account, i.e.: of the
finiteness of the speed of light it is sufficient to consider the static
eletrical polarizability $\alpha \left( 0\right) $ only, see for instance
reference \cite{Milonni}. If the electric field changes by $\delta \mathbf{E}
$\textbf{,} the interaction between the polarizable body and the electric
field will change according to $\delta V=-\mathbf{P}\left[ \mathbf{E}\right]
\cdot \delta \mathbf{E}=-\alpha \left( 0\right) \mathbf{E\cdot }\delta 
\mathbf{E}$. Therefore, if the field changes from zero to a finite value $%
\mathbf{E}$, the interaction energy is $V_{E}=-\alpha \left( 0\right) 
\mathbf{E}^{2}/2$. In the quantum version of this interaction potential we
must replace  $\mathbf{E}^{2}$ by its vacuum expectation value, $%
\left\langle \mathbf{\hat{E}}^{2}\right\rangle _{0}$. The same arguments
hold when we consider the magnetization  $\mathbf{M}$. the interaction
potential between a magnetically polarizable atom and the magnetic field is
given by $V_{M}=-\beta \left( 0\right) \mathbf{B}^{2}/2$, where $\beta \left(
0\right) $ is the static magnetic polarizability. The
correlators are given by (\ref{ECORRCASIMIR}) and (\ref{BCORRCASIMIR}) which
allow us to obtain in a straightforward way expressions for the interaction
potential energy between an electrically or magnetically polarizable atom
(placed between the plates) and the conducting plates.

Let us consider first an atom electrically polarizable placed at a distance $%
z$ from the conducting plate placed at $z=0$. The interaction potential
between is given by
\begin{equation}
V_{E}\left( z\right) =-\frac{1}{2}\alpha \left( 0\right) \left\langle 
\mathbf{\hat{E}}^{2}\left( z\right) \right\rangle _{0},
\end{equation}
where $\alpha \left( 0\right) $ is the static polarizability of the
molecule. Making use of (\ref{ECORRCASIMIR}) and (\ref{BCORRCASIMIR}) we can
evaluate $\left\langle \mathbf{\hat{E}}^{2}\left( z\right) \right\rangle _{0}
$ and using the above equation we obtain
\begin{equation}
V_{E}\left( z\right) =-\frac{\alpha \left( 0\right) \pi ^{3}}{3a^{4}}\left[
3F\left( \frac{\pi z}{a}\right) -\frac{1}{120}\right].
\end{equation}
Making use of (\ref{Fapp}) and taking the limit $a\rightarrow \infty $ we
obtain the interaction potential between a polarizable atom and a conducting
plate
\begin{equation}
V_{E}\left( z\right) =-\frac{3\alpha \left( 0\right) }{8\pi z^{4}}.
\end{equation}
If a magnetically atom or molecule is placed between conducting plates the
interaction potential will be given by
\begin{equation}
V_{M}\left( z\right) =+\frac{\beta \left( 0\right) \pi ^{3}}{3a^{4}}\left[
3F\left( \frac{\pi z}{a}\right) +\frac{1}{120}\right] ,
\end{equation}
If the atom or molecule is simultaneously electric and magntically
polarizable the interaction potential will be simply $V\left( z\right)
=V_{E}\left( z\right) +V_{M}\left( z\right) $, that is 
\begin{equation}
V\left( z\right) =-\left( \alpha \left( 0\right) -\beta \left( 0\right)
\right) \frac{\pi ^{3}}{a^{4}}F\left( \frac{\pi z}{a}\right) +\left( \alpha
\left( 0\right) +\beta \left( 0\right) \right) \frac{\pi ^{3}}{360a^{4}}.
\label{V}
\end{equation}
The single conducting plate limit $(a\rightarrow \infty )$ of (\ref{V}) is
easily obtained with the help of (\ref{Fapp}). The result is: 
\begin{equation}
V\left( z\right) \approx -\frac{3}{8\pi z^{4}}\left( \alpha \left( 0\right)
-\beta \left( 0\right) \right) ,  \label{Vapp}
\end{equation}
which is in agreement with \cite{Casimir49,Boyer69}.

The polarizable atom or molecule can be also placed between a conducting
plate at $z=0$ and a permeable one at $z=a$. In this case, making use of (%
\ref{EcorrBoyer}) e (\ref{BcorrBoyer}) a straightforward calculation leads
to the following result 
\begin{equation}
V\left( z\right) =-\left( \alpha \left( 0\right) -\beta \left( 0\right)
\right) \frac{\pi ^{3}}{a^{4}}G\left( \frac{\pi z}{a}\right) +\left( \alpha
\left( 0\right) +\beta \left( 0\right) \right) \left( -\frac{7}{8}\right) 
\frac{\pi ^{3}}{360a^{4}}.
\end{equation}
There are now two single plate limits to be considered. Near the conducting
plate at $z=0$ the potential is given by (\ref{Vapp}), but near the
perfectly permeable plate at $z=a$, the potential is repulsive and given by 
\begin{equation}
V\left( z\right) \approx +\frac{3}{8\pi \left( z-a\right) ^{4}}\left( \alpha
\left( 0\right) -\beta \left( 0\right) \right) ,
\end{equation}
where we made use of (\ref{Gapp2}). Notice that here, as mentioned before,
we dealt with a part of the interaction beteween an atom and two or one
walls. The contribution of the interaction between the electric/magnetic
dipole moment and its images was neglected. Therefore, the results refer
only to the contribution of the quantum vacuum distorted by one or two walls
to the total interaction potential. Keeping this in mind we can state that
the Casimir-Polder interaction shows certain aspects of the quantum
structure of the vacuum inbetween ans near the surfaces in question.

\section{Conclusions}

In this paper we have shown how to employ the equal time and space
electromagnetic field correlators evaluated between parallel material
surfaces to rederive results concernig the Casimir energy and pressure amd
the symmetrical traceless stress energy tensor. This is a local alternative
to the Green function technique. We have shown that for the cases we had in
mind here finite results are obtained only when we consider what happens on
both sides of the surface boundary. This consideration provided the
mechanism by which precise cancellations occurred and finite results were
obtained. This is in agreement with, for example, Ref. \cite{Deutsch79} and
should be considered as a concrete example of the behavior of quantized
fields near and on boundary surfaces. As a byproduct of these calculations
we have also analyzed the Casimir and Polder interaction between an atom and
parallel material surfaces.

\end{document}